 \documentclass{article}

\usepackage{amsmath}
\usepackage{amsfonts}
\usepackage{graphicx}
\usepackage{caption}
\usepackage{subcaption}

 \usepackage{amssymb}

\usepackage{epstopdf}
\usepackage{epsfig}
\usepackage{hyperref}

\begin{document}
\title{Higgs--Chern-Simons gravity 
models in $d=2n+1$ dimensions}
\author{{\large }
{\large Eugen Radu}$^{\dagger}$
and {\large D. H. Tchrakian}$^{\ddagger \star}$ 
\\ \\
$^{\dagger}${\small  Departamento de Matem\'atica da Universidade de Aveiro and } \\ {\small  Centre for Research and Development  in Mathematics and Applications (CIDMA),} 
\\
 {\small    Campus de Santiago, 3810-183 Aveiro, Portugal}
\\ 
$^{\ddagger}${\small School of Theoretical Physics, Dublin Institute for Advanced Studies,
10 Burlington Road, Dublin 4, Ireland}\\
$^{\star}${\small Department of Computer Science, National University of Ireland Maynooth, Maynooth, Ireland}}

\date{}
\newcommand{\dd}{\mbox{d}}
\newcommand{\tr}{\mbox{tr}}
\newcommand{\la}{\lambda}
\newcommand{\La}{\Lambda}
\newcommand{\bt}{\beta}
\newcommand{\del}{\delta}
\newcommand{\ep}{\epsilon}
\newcommand{\ta}{\theta}
\newcommand{\ka}{\kappa}
\newcommand{\f}{\phi}
\newcommand{\vf}{\varphi}
\newcommand{\F}{\Phi}
\newcommand{\al}{\alpha}
\newcommand{\ga}{\gamma}
\newcommand{\de}{\delta}
\newcommand{\si}{\sigma}
\newcommand{\Si}{\Sigma}
\newcommand{\bnabla}{\mbox{\boldmath $\nabla$}}
\newcommand{\bomega}{\mbox{\boldmath $\omega$}}
\newcommand{\bOmega}{\mbox{\boldmath $\Omega$}}
\newcommand{\bsi}{\mbox{\boldmath $\sigma$}}
\newcommand{\bchi}{\mbox{\boldmath $\chi$}}
\newcommand{\bal}{\mbox{\boldmath $\alpha$}}
\newcommand{\bpsi}{\mbox{\boldmath $\psi$}}
\newcommand{\brho}{\mbox{\boldmath $\varrho$}}
\newcommand{\beps}{\mbox{\boldmath $\varepsilon$}}
\newcommand{\bxi}{\mbox{\boldmath $\xi$}}
\newcommand{\bbeta}{\mbox{\boldmath $\beta$}}
\newcommand{\ee}{\end{equation}}
\newcommand{\eea}{\end{eqnarray}}
\newcommand{\be}{\begin{equation}}
\newcommand{\bea}{\begin{eqnarray}}
\newcommand{\z}{\zeta}

\newcommand{\ii}{\mbox{i}}
\newcommand{\e}{\mbox{e}}
\newcommand{\pa}{\partial}
\newcommand{\Om}{\Omega}
\newcommand{\om}{\omega}
\newcommand{\vep}{\varepsilon}
\newcommand{\bfph}{{\bf \phi}}
\newcommand{\lm}{\lambda}
\def\theequation{\arabic{equation}}
\renewcommand{\thefootnote}{\fnsymbol{footnote}}
\newcommand{\re}[1]{(\ref{#1})}
\newcommand{\R}{{\rm I \hspace{-0.52ex} R}}
\newcommand{\N}{{\sf N\hspace*{-1.0ex}\rule{0.15ex}%
{1.3ex}\hspace*{1.0ex}}}
\newcommand{\Q}{{\sf Q\hspace*{-1.1ex}\rule{0.15ex}%
{1.5ex}\hspace*{1.1ex}}}
\newcommand{\C}{{\sf C\hspace*{-0.9ex}\rule{0.15ex}%
{1.3ex}\hspace*{0.9ex}}}
\newcommand{\eins}{1\hspace{-0.56ex}{\rm I}}
\renewcommand{\thefootnote}{\arabic{footnote}}

\maketitle


\bigskip

\begin{abstract}
 
We consider a family of new Higgs-Chern-Simons (HCS) gravity models in $2n+1$ dimensions ($n=1,2,3$).
This provides a generalization of the (usual) gravitational Chern-Simons
(CS) gravitaties resulting from non-Abelian CS densities in all odd dimensions, which feature vector and scalar fields
in addition to the metric. The derivation
of the new HCS gravitational (HCSG) actions follows the same method as in the $usual$-CSG case resulting from the $usual$ CS
densities. The HCSG result from the HCS densities, which result through a one-step descent of the Higgs--Chern-Pontryagin (HCP),
the latter being descended from Chern-Pontryagin (CP) densities in some even dimension.
 %
	%
	A preliminary study of the solutions of these models is considered, with
	exact solutions being reported for spacetime dimensions $d=3,~5$. 
 
\end{abstract}
\medskip

\section{Introduction}

The study of the 
Chern-Simons gravities (CSG) derived from non-Abelian Chern-Simons (CS) densities
has started with the  Witten's work in
Ref.~\cite{Witten:1988hc}, dealing with the 
 $2+1$ dimensional case.
Subsequently,  Witten's results
   were extended to all odd dimensions by Chamseddine in Refs.~\cite{Chamseddine:1989nu,Chamseddine:1990gk}.
	A generic
CSG models consist of superpositions of gravitational models of all possible higher order gravities
in the given dimensions (leading to second order equations of motion), each appearing with a precise
real numerical coefficient.
These gravitational models are usually referred to as Lovelock models, which here we refer to as
$p$-Einstein gravities.
 The integer $p\geq 0$ is the power of the Riemann curvature in the Lagrangian; the
$p=0$ term being the
cosmological constant, for $p=1$ the Ricci scalar $etc$.

\medskip
The recent work \cite{Tchrakian:2017fdw,Radu:2020ytf} has proposed
 a new formulation  of the CSG systems, 
which, different from the standard case in 
\cite{Witten:1988hc}, \cite{Chamseddine:1989nu,Chamseddine:1990gk},
 allows their construction in all, $both$ odd and even dimensions. 
Following the Ref. \cite{Radu:2018fda}, let us briefly review this construction. 
As discussed there, the expression of the  new-CS densities 
is found following exactly the same method as the $usual$-CS densities in odd dimensions.
 The $usual$ CS density results
from the one-step descent of the corresponding Chern-Pontryagin (CP) density.
We recall that the CP density is a total-divergence
\[
\Om_{\rm CP}=\pa_i\Om^i\ ,\quad i=\mu,D\ ;\ \quad\mu=1,2,\dots,d\ ;\ d=D-1\ ;
\]
then the  CS density is defined as the $D$-th component of $\Om^i$, 
namely $\Om_{\rm CS}\stackrel{\rm def.}=\Om_D$.

In the proposal put forward in \cite{Tchrakian:2017fdw,Radu:2020ytf},
the role of the usual-CP density, which is defined in even dimensions only, is played by what we refer to as the
Higgs--Chern-Pontryagin (HCP) density
(see 
the Refs.~\cite{Tchrakian:2010ar,Radu:2011zy} and in Appendix A of Ref.~\cite{Tchrakian:2015pka} for a discussion of HCP models).
These are dimensional descendents of the $n^{th}$ CP density in $N=2n$ dimensions, down to residual $D$ dimensions ($D<N=2n$).
However, as a new feature, $D$ can be either odd or even. 
Also, the relics of the gauge connection on the co-dimension(s) are Higgs scalars. 
The remarkable property of the
HCP density $\Om_{\rm HCP}[A,\F]$, which is now given in terms of both the residual gauge field $A$ and the Higgs scalar $\F$, is that like the CP
density it is also a $total\ divergence$
\[
\Om_{\rm HCP}=\pa_i\Om^i\ ,\quad i=\mu,D\ ;\ \quad\mu=1,2,\dots,d\ ;\ d=D-1\ .
\]

The  corresponding new Chern-Simons density is defined by 
considering the one-step descent of the density $\Om_i$,  
as the $D$-th component of $\Om_i$, namely $\Om_{\rm HCS}\stackrel{\rm def.}=\Om_D$.
In what follows, the quantity $\Om_{\rm HCS}$ is refer to  
as the Higgs--Chern-Simons (HCS) density.
 As mentioned above, 
 such densities exist in
 both odd and even dimensions.
 Moreover, in any given dimension,
 there is an infinite family of HCS densities, following
 from the descent of a CP density in any dimension $N=2n>D$.
 A detailed discussion of these aspects is given in
Refs.~\cite{Tchrakian:2010ar,Radu:2011zy,Tchrakian:2015pka}.
Note that a similar definition for the HCS density was proposed in Ref. \cite{Szabo:2014zua}
but only in odd dimensions 
and with the Higgs scalar being a complex column, not suited to the application here.

\medskip

With this definition of 
  the  HCS  densities,
	the construction of the corresponding
	 gravitational theories
	is done
	 in the same spirit
as in \cite{Witten:1988hc,Chamseddine:1989nu,Chamseddine:1990gk}.
	%
In any given dimension, there is an infinite family of such theories, each resulting from the infinite family of HCS densities.
Working in $d=D-1$ dimensions, the gauge group is
chosen to be $SO(d)$,
while 
the Higgs multiplet is chosen to be a
$D$-component $isovector$ of $SO(D)$~\footnote{These choices coincide
with the representations that yield monopoles on $\R^d$, as described in \cite{Tchrakian:2010ar}. }.
%
The central point in the  construction  of both CS and  HCS gravity  models is the identification of 
the non-Abelian (nA) $SO(D)$ connection in $d=D-1$
dimensions~\footnote{Note that no choice for the signature of the space is make at this stage.},
with the spin-connection $\om_\mu^{ab}$ and the $Vielbein$ $e_\mu^a$, ($\mu=1,2,3$;\ a=1,2,3).
Following  the
prescription in \cite{Witten:1988hc,Chamseddine:1989nu,Chamseddine:1990gk},
 we define
\bea
A_{\mu}&=&-\frac12\,\om_{\mu}^{ab}\,\ga_{ab}+\ka\,e_{\mu}^{a}\,\ga_{aD}\quad\Rightarrow\quad
F_{\mu\nu}=-\frac12\left(R_{\mu\nu}^{ab}-\ka^2\,e_{[\mu}^{a}\,e_{\nu]}^{b}\right)\ga_{ab}\,,\label{oddd}
\eea
$(\ga^{ab},\ga^{aD})$ being the Dirac gamma matrices used in the representation of the algebra of $SO(D)$. 
Note the presence of the constant $\ka$ in the above expression (with  
dimensions $L^{-1}$), compensating the difference of the dimensions of the spin-connection and the $Dreibein$. In \re{oddd},
$$R_{\mu\nu}^{ab}=\pa_{[\mu}\om_{\nu]}^{ab}+(\om_{[\mu}\om_{\nu]})^{ab}$$ is the Riemann curvature.

In the  HCS case, in addition to \re{oddd}, we supplement (\ref{oddd}) 
with the prescription for the Higgs scalar $\F$,
\bea
2^{-1}\F=(\f^{a}\,\ga_{a,D+1}+\psi 
\,\ga_{D,D+1})\  &\Rightarrow& \ 
2^{-1}D_{\mu}\F=(D_{\mu}\f^{a}-\ka\,e_\mu^a\,\psi)\ga_{a,D+1}+(\pa_{\mu}\psi+\ka\,e_\mu^a\,\f^a)\ga_{D,D+1}
\label{higgsevenD2}
\eea
which clearly displays the $iso$-$D$-$vector$ 
$(\f^a,\f^D)$, 
that is split into the $D\ component$ frame-vector
field $\f^a$ and the scalar field 
$\psi \equiv\f^D$. 
Also, we define the 
 covariant derivative $D_\mu\F$ of the Higgs scalar
\be
\label{gravcov}
D_\mu\f^a=\pa_\mu\f^a+\om_\mu^{ab}\f^b~.
\ee
Note also that  $\f^a$ is a vector field 
(with $\f_\mu=e_\mu^a\f^a$ 
  in a coordinate frame)
 which, however, has rather unusual dynamics as will be seen below. 
As such, $\f^a$ is neither a gauge (masless or massive) field; it has rather a geometric content. 
 
In fact, 
one remarks that the fields $(\f^a,\psi)$ are not usual matter fields like gauge fields or Higgs scalar in the Standard Model of particle physics.
 In the latter
cases, the covariant derivatives are not defined by the (gravitational) spin-connection, while here they are, as
seen in \re{higgsevenD2}.   
Thus, in this sense $(\f^a,\psi)$ are like spinor fields (although their action look different). 
As such,  theories like the one proposed here can support solutions with torsion,
a possibility, which, however, is not explored in this work.
However, the analogy with spinors is incomplete, since 
 the fields $(\f^a,\psi)$ are rather 'gravitational coordinates', as they originate from the Higgs field
$\F$ of the nA gauge theory.
%
Thus, as seen from \re{higgsevenD2}, $(\f^a,\psi)$
are on the same footing as the $usual$ 'gravitational coordinates' $(\om_\mu^{ab},e_\mu^a)$.  
In fact, this provides 
the main physical  motivation for their study, since such models can be seen
as extensions of the usual CSG's (reducing to them in the limit of vanishing  $(\f^a,\psi)$).
Therefore it is interesting to see what are the new features 
introduced by extending the CSG's to allow for nonzero $(\f^a,\psi)$.
Moreover, finding how the BTZ-like CSG's  black hole solutions in Ref.\cite{Cai:1998vy}
are deformed by the $(\f^a,\psi)$-fields is an interesting mathematical problem in itself.
 
 The gravitational models resulting from the Higgs-CS (HCS) densities via \re{oddd}-\re{higgsevenD2} are
referred to as HCS gravities~\cite{Tchrakian:2017fdw} (HCSG). 
In this report we  restrict  our  
study to the
 simplest  HCSG models in $2n+1$ dimensions, namely to the HCSG models resulting from the HCS density descended from
the HCP densities in $6$, $8$ and $10$ dimensions. 
These models are extensions of the
$usual$ Chern-Simons gravities~\cite{Witten:1988hc,Chamseddine:1989nu,Chamseddine:1990gk},
possessing an additional sector in terms of $(\f^a,\psi)$.

 This paper is organized as follows. 
In the next Section, we present
the explicit form of the resulting HCSG Langrangians for $d=3,5,7$,
 the connection with the usual CSG being also reviewed.
A preliminary investigation of the simplest solutions in $d=3,~5$ dimensions is  considered in Section 3, 
extending the study (for  the $n=1$ case) in  Ref.~\cite{Radu:2018fda}. 
The concluding remarks are presented in  Section 4.

\section{HCSG models in $2n+1$ dimensions, $n=1,2,3$}
%
\subsection{General expressions}
%
The Higgs--Chern-Simons densities (HCS) considered here are the ``simplest'' examples in $2+1,\ 4+1,\ 6+1$ dimensions.
By simplest we mean that the Higgs--Chern-Simons (HCS) density employed to construct the HCS gravity (HCSG), is the one resulting from the
``simplest'' Higgs--Chern-Pontryagin (HCP) density, which is defined in one dimension higher, namely in $four$, $six$ and $eight$ dimensions respectively. Now in $any$ dimension,
HCP densities can be constructed as dimensional descendants of a CP density in $2n>4$ dimensions, hence it is reasonable to describe the
``simplest'' cases here to be the HCP densities in $4,\ 6,\ 8$ dimensions, that descend from the CP densities in $2n=6,\ 8,\ 10$ dimensions respectively, $i.e.$, those descended by the minimal (nontrivial) number of dimensions, namely by $two$ dimensions.

Since like the CP density, the HCP density is a $total\ divergence$, then the corresponding HCS density results from the usual $one$-$step$ descent, which
in the cases at hand are those from $4,\ 6,\ 8$ to $3,\ 5,\ 7$ dimensions, with the corresponding expressions
%
\bea
\Omega^{(3,6)}_{\rm HCS}&=&
\eta^2\Omega_{\rm CS}^{(3)}+\vep^{\mu\nu\la}\mbox{Tr}\,\ga_5D_{\la}\F(\F F_{\mu\nu}+F_{\mu\nu}\F)\,,
\label{HCS36}
\\
\Omega^{(5,8)}_{\rm HCS}&=&\eta^2\Om_{\rm CS}^{(5)}+\vep^{\mu\nu\rho\si\la}\,\mbox{Tr}\,\ga_7
D_{\la}\F\left(\F F_{\mu\nu}F_{\rho\si}+F_{\mu\nu}\F F_{\rho\si}+F_{\mu\nu}F_{\rho\si}\F\right),
\label{HCS58}
\\
\Omega^{(7,10)}_{\rm HCS}&=&\eta^2\Om_{\rm CS}^{(7)}
                             +\vep^{\mu\nu\rho\si\tau\la\ka}\,\mbox{Tr}\,\ga_9D_{\ka}\F(\F F_{\mu\nu}F_{\rho\si}F_{\tau\la}+F_{\mu\nu}\F F_{\rho\si}F_{\tau\la}\nonumber\\
&&\hspace{50mm}+F_{\mu\nu}F_{\rho\si}\F F_{\tau\la}+F_{\mu\nu}F_{\rho\si}F_{\tau\la}\F)\,.\label{HCS710}
\eea

Let us remark that 
the HCS densities \re  {HCS36}, \re{HCS58} and \re{HCS710} are the ``simplest'' examples in these dimensions, arising from the descents of the Chern-Pontryagin densities in $6$, $8$ and $10$ dimensions respectively. It may be interesting  to display  also HCS densities
arising from CP densities in higher dimensions. 
To this end we consider the HCS density in $3$ dimensions resulting from the descent from the CP in $8$ dimensions
\bea
\Omega^{(3,8)}_{\rm HCS}&=&2\eta^4\Omega_{\rm CS}^{(3)}+(2\eta^2-|\f^a|^2-\f^2)\,\vep^{\mu\nu\la}\,\mbox{Tr}\,\ga_5\,
D_{\la}\F(\F F_{\mu\nu}+F_{\mu\nu}\F)
\label{HCS38}                 
\eea

\medskip
The leading term $\Om_{\rm CS}^{(3)}$ in \re{HCS36} and \re{HCS38}, and the leading terms $\Om_{\rm CS}^{(5)}$ and $\Om_{\rm CS}^{(7)}$ in \re{HCS58} and\re{HCS710} are the usual CS densities for
the $SO(D),\ D=4,\ 6,\ 8$ gauge connection,
\bea
\Omega_{\rm CS}^{(3)}&=&\vep^{\la\mu\nu}\mbox{Tr}\,\ga_5
A_{\la}\left[F_{\mu\nu}-\frac23A_{\mu}A_{\nu}\right],
\label{CS32}\\
\Omega_{\rm CS}^{(5)}&=&\vep^{\la\mu\nu\rho\si}\mbox{Tr}\,\ga_7
A_{\la}\left[F_{\mu\nu}F_{\rho\si}-F_{\mu\nu}A_{\rho}A_{\si}+
\frac25A_{\mu}A_{\nu}A_{\rho}A_{\si}\right],
\label{CS52}
\\
\Omega_{\rm CS}^{(7)}&=&\vep^{\la\mu\nu\rho\si\tau\ka}
\mbox{Tr}\,\ga_9A_{\la}\bigg[F_{\mu\nu}F_{\rho\si}F_{\tau\ka}
-\frac45F_{\mu\nu}F_{\rho\si}A_{\tau}A_{\ka}-\frac25
F_{\mu\nu}A_{\rho}A_{\si}F_{\tau\ka}\nonumber\\
&&\qquad\qquad\qquad\qquad\qquad\qquad
+\frac45F_{\mu\nu}A_{\rho}A_{\si}A_{\tau}A_{\ka}-\frac{8}{35}
A_{\mu}A_{\nu}A_{\rho}A_{\si}A_{\tau}A_{\ka}\bigg]\,.\label{CS72}
\eea
In \re{HCS36}-\re{HCS710}, the Higgs scalar $\F$, the gauge connection
and
the constant $\eta$  have the dimensions of $L^{-1}$.

Applying the prescriptions \re{oddd} and \re{higgsevenD2} to \re{HCS36}, \re{HCS58}, \re{HCS710} and \re{HCS38}
yield the required HCS gravitational (HCSG) models.  
To express these compactly, we adopt the abbreviated notation
\bea
{\bar R}_{\mu\nu}^{ab}&=&R_{\mu\nu}^{ab}-\ka^2\,e_{[\mu}^ae_{\nu]}^b,
\label{abbr1}
\eea
together with
\bea
\f_\mu^a&=&D_\mu\f^a-\ka\,e_\mu^a\psi,
\label{abbr2}
\\
\psi_\mu&=&\pa_\mu\psi +\ka\,e_\mu^a\f^a \,,\label{abbr3}                          
\eea
where $R_{\mu\nu}^{ab}$ is the Riemann curvature and $D_\mu\f^a$ is the covariant derivative \re{gravcov},
of the frame-vector field $\f^a$.

In terms of which the HCSG Lagrangians in $d=3$, $5$ and $7$ dimensions, resulting from the HCS densities \re{HCS36}, \re{HCS58}, \re{HCS710}, are
\bea
{\cal L}_{\rm HCSG}^{(3)}&=&\eta^2\ka\,
{\cal L}_{\rm CSG}^{(3)}+\vep^{\mu\nu\la}\vep_{abc}{\bar R}_{\mu\nu}^{ab}
(\psi\,\f_\la^c-\f^c\psi_\la),
\label{HCSG36}
\\
{\cal L}_{\rm HCSG}^{(5)}&=&\eta^2\ka\,{\cal L}_{\rm CSG}^{(5)}
-\frac34\vep^{\mu\nu\rho\si\la}\vep_{abcde}{\bar R}_{\mu\nu}^{ab}{\bar R}_{\rho\si}^{cd}(\psi\,\f_\la^c-\f^c\psi_\la),\label{HCSG58}
\\
{\cal L}_{\rm HCSG}^{(7)}&=&\eta^2\ka\,{\cal L}_{\rm CSG}^{(7)}
+2\vep^{\mu\nu\rho\si\tau\la\ka}\vep_{abcdefg}{\bar R}_{\mu\nu}^{ab}{\bar R}_{\rho\si}^{cd}{\bar R}_{\tau\la}^{ef}
(\psi\,\f_\la^c-\f^c\psi_\la),
\label{HCSG710}
\eea
while the gravitational model arising from \re{HCS38} is
\be
{\cal L}_{\rm HCSG}^{(3,8)}=2\eta^4\ka\,{\cal L}_{\rm CSG}^{(3)}+\vep^{\mu\nu\la}\vep_{abc}(2\eta^2-|\f^d|^2-\psi^2)
\,{\bar R}_{\mu\nu}^{ab}
(\psi\,\f_\la^c-\f^c\psi_\la)
\,.
\label{HCSG38}
\ee
In \re{HCSG36}-\re{HCSG710}, 
${\cal L}_{\rm CSG}^{(3)}$, ${\cal L}_{\rm CSG}^{(5)}$ and ${\cal L}_{\rm CSG}^{(7)}$ are the usual Chern-Simons gravities (CSG)
\bea
{\cal L}_{\rm CSG}^{(3)}&=&-\vep^{\mu\nu\la}\vep_{abc}\left(R_{\mu\nu}^{ab}-\frac23\,\ka^2e_{\mu}^ae_{\nu}^b\right)e_{\la}^c,
\label{3csg}
\\
{\cal L}_{\rm CSG}^{(5)}&=&\vep^{\mu\nu\rho\si\la}\vep_{abcde}\left(\frac34\,R_{\mu\nu}^{ab}\,R_{\rho\si}^{cd}-\ka^2\,R_{\mu\nu}^{ab}\,e_{\rho}^ce_{\si}^d
+\frac35\,\ka^4e_{\mu}^ae_{\nu}^be_{\rho}^ce_{\si}^d\right)e_{\la}^e,
\label{5csg}
\\
{\cal L}_{\rm CSG}^{(7)}&=&-\vep^{\mu\nu\rho\si\tau\ka\la}\vep_{abcdefg}\bigg(\frac18\,R_{\mu\nu}^{ab}\,R_{\rho\si}^{cd}\,R_{\tau\ka}^{ef}
-\frac14\,\ka^2\,R_{\mu\nu}^{ab}\,R_{\rho\si}^{cd}\,e_{\tau}^ee_{\ka}^f
\label{7csg}
\\
&&
\qquad\qquad\qquad\qquad+\frac{3}{10}\,\ka^4\,R_{\mu\nu}^{ab}\,e_{\rho}^ce_{\si}^de_{\tau}^ee_{\ka}^f
-\frac17\,\ka^6e_{\mu}^ae_{\nu}^be_{\rho}^ce_{\si}^de_{\tau}^ee_{\ka}^f\bigg)e_{\la}^g~.
\nonumber
\eea
It is easy to express the HCSG Lagranian for all $n$ by extrapolation of \re{HCSG36}-\re{HCSG710}.
 
Given the models \re{HCSG36}-\re{HCSG710} ,
the corresponding equations of motion are found by taking the variation of the action 
$w.r.t.$ the vielbein $e_\lambda^a$, together with $(\f^a,\psi)$.
However, these equations
have a simple enough expression for  $d=3$ only, see  Ref.~\cite{Radu:2018fda}.

\subsection{The general CSG Lagrangians and the connection with the Einstein-Lovelock hierarchy}

Following the previous study \cite{Radu:2018fda},
we consider a spacetime with  Minkowskian signature, and  replace
\be
\label{replace}
\ka\to i\ka\ ,\quad h\to -ih\,.
\ee
in the Lagrangians \re{HCSG36}-\re{HCSG710}.
With this choice, setting $\phi=\phi^a=0$
results in (pure gravity) CS Lagrangian in $d=2n+1$ dimensions, with a negative cosmological constant
$\Lambda$. 
Also, note that one can $\eta=1$ without any loss of generality,
a choice we employ for the rest of this work.

The  CS Lagrangian in $d=2n+1$ dimensions
can be viewed as a particular case of a
  generic model consisting in a 
  superposition of all allowed Einstein-Lovelock terms
	in that dimension, with
\begin{eqnarray}
\label{Lgen}
 {\cal L}_{\rm CSG}^{(2n+1)}
=\sum_{p=0}^{n}
\alpha_{(p)}{L}_{(p)}, 
\end{eqnarray}
  with the following definition for  the $p-$th term in the Lovelock hierarchy. 
\begin{eqnarray}
\label{Lp-my}
{  L}_{(p)}=\frac{p!}{2^p}
\delta^{\mu_1}_{[\rho_1}\cdots \delta^{\mu_p}_{\rho_p]}
R_{\mu_1\nu_1}^{\phantom{\mu_1}\phantom{\nu_1}\rho_1\sigma_1}\cdots R_{\mu_p\nu_p}^{\phantom{\mu_p}\phantom{\nu_p}\rho_p\sigma_p}.
\end{eqnarray}
The normalization of each term in (\ref{Lp-my}) has been chosen 
to make contact with the usual conventions in the literature on Lovelock gravities solutions.
As such, 
 ${  L}_{(p)}=R^{p}+\dots$,
the first terms (up to $d=7$) being 
\begin{eqnarray}
&&{  L}_{(0)} = 1,
~~{ L}_{(1)} = R,
~~
{  L}_{(2)} = R^2-4R_{\mu\nu}R^{\mu\nu}+R_{\mu\nu\rho\sigma}R^{\mu\nu\rho\sigma},
\\
&&
{  L}_{(3)} = R^3   -12RR_{\mu \nu } R^{\mu \nu } + 16R_{\mu \nu }R^{\mu }_{\phantom{\mu } \rho }R^{\nu \rho }+ 24 R_{\mu \nu }R_{\rho \sigma }R^{\mu \rho \nu \sigma }+ 3RR_{\mu \nu \rho \sigma } R^{\mu \nu \rho \sigma } 
\nonumber 
\\
&&{~~~~~~~~~}
-24R_{\mu \nu }R^\mu _{\phantom{\mu } \rho \sigma \kappa }
 R^{\nu \rho \sigma \kappa  }+ 4 R_{\mu \nu \rho \sigma }R^{\mu \nu \eta \zeta } 
R^{\rho \sigma }_{\phantom{\rho \sigma } \eta \zeta }
-8R_{\mu \rho \nu \sigma } R^{\mu  \phantom{\eta } \nu }_{\phantom{\mu } \eta 
\phantom{\nu } \zeta } R^{\rho  \eta  \sigma  \zeta }.
\end{eqnarray}
Also, 
to make contact with the usual GR conventions, we take 
\begin{eqnarray}
\label{01}
 \alpha_{(0)}=-2\Lambda,~~\alpha_{(1)}=1.
\end{eqnarray}
In general, the coefficients  $\alpha_{(p)}$ are arbitrary.
However, they are fixed in a CGS model, with the general expression
 \begin{eqnarray}
\label{expr}
 \alpha_{(p)}=(-1)^{p+1}
\frac{1}{\Lambda^{p-1}}
\frac{(d-2)^p}{2^{p-1}p! (d-2p)}\frac{(d-2p)!!}{(d-2)!!},
\end{eqnarray}
 where $\Lambda =-2 (d-2) \kappa^2$.


\section{The solutions}

Given the above models  \re{HCSG36}-\re{HCSG710},
it is interesting to inquire which are the  simplest solutions
with nonvanishing fields $(\phi^a,\psi)$.
In what follows, we study this question for the first two dimensions $d=3,5$, 
and contrast the results. 

\subsection{The $d=3$ case}

We consider a static line element with
\begin{eqnarray}
\label{d31}
ds^2=\frac{dr^2}{N(r)}+r^2  d \varphi^2-N(r) e^{-2\delta(r)} dt^2
\end{eqnarray}
where $r,t$ are the radial and time coordinates, respectively, while 
$ \varphi$ is the azimuthal  coordinate.
%
Working in a coordinate basis,  a consistent Ansaz for the fields $(\f^a,\psi)$ reads
\begin{eqnarray}
\label{gen-matter}
\phi=f(r) dr+g(r) dt,~~~\psi\equiv h(r).
\end{eqnarray}

Then a straightforward computation  leads to
 the following exact solution of the full set of equations of motion:
\begin{eqnarray}
\label{ex31}
&&
N(r)=\kappa^2 r^2-n_0,~~\delta(r)=0,
\\
&&
\label{ex32}
f(r)=\frac{c_0}{\kappa}+\frac{c_1}{N},~~g(r)=\sqrt{c_1^2+c_2 N(r)},~~\psi \equiv h(r)=c_0 r,
\end{eqnarray}
with $n_0,c_0,c_1,c_2$
arbitrary constants.
This provides a generalization of the solution reported in 
\cite{Radu:2018fda}, which is recovered\footnote{Note that the solution in
 \cite{Radu:2018fda} is expressed in a $dreibein$
basis with $e_r=1/\sqrt{N}$, $e_\varphi=r d\varphi$, $e_t=\sqrt{N}dt$.} 
for $c_1=c_2=0$.

One can see that the choice
 $n_0=-1$ corresponds to a globally AdS$_3$
geometry, while
for $n_0>0$ the BTZ  black hole (BH) geometry \cite{Banados:1992wn} is recovered. 
In both cases, the fields $(\phi^a,\psi)$ do
not backreact on the spacetime (thus their contribution to the $r.h.s.$
of the Einstein equations with negative cosmological constant vanishes\footnote{The analogy of these solutions with
 self-dual Yang-Mills instantons in a  curved space geometry
	\cite{Charap:1977ww}, 
	\cite{Brihaye:2006bk},
	\cite{Radu:2007az}, 
was noticed in Ref.~\cite{Radu:2018fda}. There, we dubbed these closed form solutions as {\it effectively vacuum confiurations}.
	Another interesting analogy is provided by the  'stealth' BH solutions in various
		alternative models of gravity 
		(see $e.g.$ 
		\cite{Bernardo:2019yxp},
		\cite{Takahashi:2020hso}
		and references therein). 
		Such configurations feature a nontrivial scalar field, while the geometry is still that of the (vacuum)
general relativity solutions.
}).
However,   $(\phi^a,\psi)$ possess a nonstandard behaviour 
($e.g.$ both $\psi$
and $|\vec\f|^2$ diverge as $r\to \infty$).
Moreover, the Ref. \cite{Radu:2018fda} has given arguments that 
(at least in the
$c_1=c_2$ limit), 
the solution (\ref{ex31}),  (\ref{ex32}) appears to be unique.
Since the discussion here in $d=3$ is similar to that in the $d=5$ case,
 we restrict to the discussion of the latter, below.

Finally, we mention the existence of a generalization of the solution in 
Ref. \cite{Radu:2018fda} for a spinning BTZ background.
The line element in this case reads
\begin{eqnarray}
\label{1}
ds^2=\frac{dr^2}{N(r)}+r^2(d\varphi-W(r)dt)^2 -N(r)dt^2,~~{\rm where~~}N(r)=\kappa^2 r^2 -n_0+\frac{J^2}{r^2},~~W(r)=\frac{J}{r^2},
\end{eqnarray}	 
 with $J$ the angular momentum.
While the expression of the scalar field remains the same, the function is more complicated, 
\begin{eqnarray}
\label{3}
h(r)=c_0 r,~~
f(r)=\frac{c_0}{\kappa}
\left( 
1+\frac{J}{r \sqrt{N}}
\right).
\end{eqnarray}	
One can easily see that all unusual features noticed in the static case
for  functions $f$, $h$ are present also in this case.

\subsection{The $d=5$ case}

 \subsubsection{An exact solution}

The metric Ansatz here is more complicated, the $S^1$ direction in the $d=3$ line-element
(\ref{d31})
being replaced with a surface of constant curvature.
As such, we consider a general line-element
\begin{eqnarray}
\label{gen-metric}
ds^2=\frac{dr^2}{N(r)}+r^2 d \Sigma^2_{k,3} -N(r) e^{-2\delta(r)} N(r) dt^2
\end{eqnarray}
with $k=0,\pm 1 $,
while the three--dimensional metric $d\Sigma^2_{k,3}$ is
\begin{equation}
d\Sigma^2_{k,3} =\left\{ \begin{array}{ll}
\vphantom{\sum_{i=1}^{3}}
 d\Omega^2_3 & {\rm for}\; k = +1\\
\sum_{i=1}^{3} dx_i^2&{\rm for}\; k = 0 \\
\vphantom{\sum_{i=1}^{d-3}}
 d\Xi^2_3 &{\rm for}\; k = -1\ .
\end{array} \right.
\end{equation}
In the above relation, $d\Omega^2_3$ is the unit metric on $S^3$ $(k=1)$;
for $k=0$ we have a flat three-surface;
while for $k=-1$, one considers 
a the three--dimensional hyperbolic space, whose unit 
metric  $d\Xi_3^2$ can be obtained by analytic continuation of 
that on $S^3$.

The ansatz for the fields $(\phi^a,\psi)$ is still given by
(\ref{gen-matter}).
Within this framework,
the following  closed-form solution of 
 the field equations has been found
\begin{eqnarray}
\label{sol1}
N(r)=\kappa^2 r^2+k ,~~\delta(r)=0,
\end{eqnarray}
together with
\begin{eqnarray} 
\label{sol2}
h(r)=\frac{c_0}{r},~~
f(r)= \frac{c_0}{\kappa r^2} \left( -1+(k^2+k-1)\sqrt{1+\frac{3\kappa^2 r^2}{N(r)}}  \right),~~
g(r)=0~,
\end{eqnarray}
with $c_0$ an arbitrary constant.
The corresponding line elements,
as implied by (\ref{sol1}) are well known, 
corresponding to three 
different parametrization of AdS$_5$ spacetime.
Although they possess the same (maximal) number of Killing symmetries,
they present different global properties, the case $k=0$
corresponding to a Poincar\'e patch and the gloabbly AdS$_5$ spacetime
being found for $k=1$
(see $e.g.$
the discussion in Ref. \cite{Emparan:1999pm}).
 
As with the $d=3$ case above, the fields $(\phi^a,\psi)$
do not backreact on the spacetime geometry,
$i.e.$ their  effective  energy-momentum tensor vanishes again.
However, while  this time  both $\psi$
and $|\vec\f|^2$ are finite as $r\to \infty$,
they diverge at the minimal value of $r$ (which is $r=0$ for $k=0,1$
and $r=1/\kappa$ for $k=-1$).

Finally, let us remark that the expressions (\ref{sol2}) for  $(\phi^a,\psi)$
are also compatible with a different 
  background, described by the line-element
	(where
 $k=0,\pm 1 $)
\begin{eqnarray}
 \label{sup1}
ds^2=\frac{dr^2}{N(r)}+r^2 d \Sigma^2_{k,2}+r^2 dz^2 -N(r) dt^2,~~{\rm with }~~
N(r)=\kappa^2 r^2 +\frac{k}{3}.
\end{eqnarray}
where $-\infty<z<\infty$ while
$d\Sigma^2_{k,2}$ the metric  on a two--dimensional surface
  of constant $2k$-curvature.
	For $k=0$ the Poincar\'e patch of AdS$_5$ spacetime is recovered;
	the case $k=1$ corresponds to a vortex-type geometry, while a black string is recovered for $k=-1$.
As with the line-element (\ref{gen-metric}),
this is a solution of the equations of motion also for $f=g=h=0$,
whose existence has been noticed in Ref. \cite{Brihaye:2013vsa}.

\subsubsection{No backreacting solutions 
on a fixed black hole background
}
 
For $\phi^a=\psi=0$,
the CSG equations of motion 
(with the line-element (\ref{gen-metric}))
 possess the exact solution
\begin{eqnarray}
\label{sol1n}
N(r)=\kappa^2 r^2- n_0,~~ \delta(r)=0,
\end{eqnarray}
 with $n_0$ an arbitrary constant.
The AdS$_5$ line-element discussed above is a particular case here,
corresponding to the choice  $n_0=-k$.
However, $n_0>0$ leads to a BTZ-like BH geometry \cite{Cai:1998vy},
with an horizon located at $\sqrt{n_0}/\kappa$.
Moreover, one can show that
the same expression of the metric functions
 $N(r)$, $\delta(r)$ solves the CS gravity equation in all $d=2n+1$ dimensions
(for a choice of the line element similar to (\ref{gen-metric}), 
$d\Sigma^2_{k,3}$ being replaced with its higher dimensional generalization),
 see
the discussion in the recent work
 \cite{Tchrakian:2019iem} and the references there.

\medskip

 A major difference between the 
$d=3$ and $d=5$
cases appear to be that for $d=5$  we could not extend 
(\ref{sol1}) to include the case of a BH solution, $n_0\neq k$.
However, 
  this results follows directly from the structure of the field equations,
	which is different for $d=3$ and $d=5$.
Let us assume that the geometry (\ref{gen-metric}) 
with $(N,\delta)$ as given by (\ref{sol1n}))
is a solution of the $d=5$ model.
Then the equations for the  functions   $f,g,h$
take the simple form  (note the relations below are for $k=1$, only;
however, a similar result is found also for $k=0,-1$):
\begin{eqnarray}
\nonumber
&&
h'-\kappa f-\frac{\kappa}{2}(1-\frac{1}{N^2}) g=0,
\\
\label{sup1s}
&&
h'-\frac{1}{2}\kappa (1-N^2)f-\frac{n_0+\kappa^2 r^2 N^2}{2r N}h-\kappa g=0,
\\
\nonumber
&&
g'-N^2 f'+\frac{\kappa^2 r^2 N^2-2N-3n_0}{2rN}g
-\frac{2n_0+3N}{r}f+3\kappa N h=0.
\end{eqnarray}
It follows directly that both $f$ and $g$
can be expressed in term of $h$, with
\begin{eqnarray}
\label{supw}
f=\frac{1}{2} 
(-1+\frac{1}{N^2}) g+\frac{h'}{\kappa},
\end{eqnarray}
and
\begin{eqnarray}
\label{supws}
g=\frac{2N^2}{\kappa (1+N^2)}h'
-\frac{2N(n_0+\kappa^2 r^2 N^2)}{\kappa r (1+N^2)^2} h=0~.
\end{eqnarray}
The scalar $ h$
is a solution of an equation of the form
\begin{eqnarray}
\label{sup3}
 (1+n_0)h U(r)=0,~~{\rm with}~~U(r)=\sum_{k=0}^{5}c_k(\kappa,n_0)r^{2k},
\end{eqnarray}
the explicit form of $c_k$ being irrelevant.
As such, for $h\neq 0$ the only choice is $n_0=-1$,
$i.e.$ a globally AdS$_5$ spacetime.
Then the matter fields equations are satisfied,
the functions $f$ and $g$ being fixed by $h$.
The expression of the scalar $h$ is found by 
imposing the gravity equations to be satisfied for the above choice of the geometry,
which result in the solution (\ref{sol2}).
 
A similar computation for $d=3$ leads again to a set of three equations for $(\phi^a,\psi)$,
which again reduce to a single equation for $h(r)$.
However, this time
this equation is multiplied with a factor 
$N'-2\kappa^2 r$.
Therefore the choice $N=\kappa^2 r^2-n_0$, with $n_0$ arbitrary,
is now allowed. 
The the solution  (\ref{ex32}) 
is recovered when imposing the Einstein equation to be also satisfied.

\medskip 
Returning to the $d=5$ case,
one may ask if a more general solution  exists, with the functions    
$(\phi^a,\psi)$ backreacting on the spacetime metric and
being finite everywhere.
The answer seems to be negative,
although we do not have a definite proof.
An indication comes from from our attempt to construct a numerical solution.
Here one starts by noticing that,
starting with the general framework 
(\ref{gen-matter}), (\ref{gen-metric}),
the function $f$ can be eliminated (as found from the field equations), 
with
 \begin{eqnarray}
f(r)=\frac{h'(r)}{\kappa},
\end{eqnarray}
where we assume that globally AdS spacetime is $not$ a solution.
Also, one can prove that $g=0$ is a consistent truncation of the model.
As such, we are left with three ordinary differential  equations for the functions 
$h$, $N$ and $\delta$.
Restricting to the most interesting $k=1$ case 
($i.e.$ a globally AdS$_5$ background),
we have attempted to construct
 deformations of the line element
(\ref{sol1}), with a regular origin and usual AdS$_5$ asymptotics, which
would represent
 particle-like solitonic configurations.
In our approach,
 we assume that the
small-$r$ solution 
possesses a power series expansion, with 
\begin{eqnarray}
 \label{ssc}
N(r)= \sum_{k\geq 0}n_{(k)} r^k,~~~
 \delta(r)= \sum_{k\geq 0}\delta_{(k)}r^k~~~{\rm and}~~~
 h(r)=\sum_{k\geq 0}h_{(k)} r^k,~~
\end{eqnarray}
where $n_{(k)}$, $\delta_{(k)}$ and $h_{(k)}$ are real numbers (and $n_0=1$)
subject to a tower of algebraic conditions, as implied by the field equations.
Starting with the above small-$r$ expansion, we have integrated   the HCS equations of motion,
searching for solutions with 
$N(r)\to \kappa^2 r^2+const. $,   
$\delta\to 0$ and 
$h(r)\to h_0$
(with $h_0$ a constant),
as $r\to \infty$.
However, we have failed to find any numerical indication for the existence of such configurations,
the solutions possessing 
a pathological behavious for 
 any considered set of initial conditions at $r=0$,
typically with a the occurence of a divergence at  some finite $r$.

A similar results holds also for BH configurations, 
in which case
  we assume the existence of an horizon at some $r=r_H>0$,
with $N(r_H)=0$,
 while $\delta (r_H)$
 and $h(r_H)$ nonzero and finite.
 
Finally, let us remark that,
although a definite proof is  missing, the above (numerical) results follow
 the spirit of the 'no hair' theorems 
\cite{Bekenstein:1996pn},
\cite{Volkov:1998cc},
\cite{Herdeiro:2015waa},
as expressed in the conjecture that there are no BH solutions with matter fields 
 that do not possess (asymptotically)
measured quantities subject to  a Gauss Law.

\section{Conclusions}

Chern-Simons gravity (CSG) 
models in $d=2n+1$ dimensions were extensively studied in the literature, starting with 
Witten's work for $d=3$  \cite{Witten:1988hc},
 where the gravitational model is described by the Einstein-Hilbert  Lagrangian with a  cosmological constant. 
In the  $d>3$ case, such systems consist of specific superpositions of gravitational
Lagrangians featururing all possible powers of the Riemann curvature of in the given dimension, each appearing with a
precise numerical coefficient.
The main purpose of this paper was to propose a generalization of the CSG model,
with a Lagrangian,
which in addition to the (standard) CSG Lagrangian, 
  features new terms described by a frame-vector
   field $\f^a$ and a scalar field $\psi$. Like the CSG, which result from the non-Abelian (nA) Chern-Simons (CS) densities,
these new Lagrangians result from a new class of CS densities which in addition to the nA gauge field, feature an algebra-valued Higgs scalar.
Like the usual nA CS densities, which result from the usual Chern-Pontryagin (CP) densities, 
these new CS densities are constructed in the same way, but now from the dimensional descendents of the CP densities which feature the Higgs scalar.
The latter are referred to as Higgs--Chern-Pontryagin (HCS)~\cite{Tchrakian:2010ar,Radu:2011zy,Tchrakian:2015pka}
densities, and are the building blocks for the generalised CSG's, namely the HCSG's~\cite{Tchrakian:2017fdw,Radu:2020ytf} studied here.

It should be noted at this stage that the construction of HCSG's is not confined to odd dimensions only, since the HCS from which they are constructed
are defined is both odd $and$ even dimensions.
 The main reason we have restricted our attention to odd dimensions in these preliminary investigations is
that only in odd dimensions there exist CSG's, which can provide a background for the new gravitational field configurations.
In even dimensional spacetimes,  the  HCSG models,  
as typified by the $3+1$ dimensional 
examples in Refs.~\cite{Tchrakian:2017fdw,Radu:2020ytf},
also consist  of frame-vector and scalar fields $(\f^a,\psi)$ interacting with the gravitational $Vielbein\ e_\mu^a$ (or the metric).
These Lagrangians are invariant under gravitational gauge transformations;
however, different from the odd dimensional case in this work,  they do not feature (gauge-variant, pure gravity) CSG terms,
their action mixing the contribution
of  $(\f^a,\psi)$, $ e_\mu^a$ fields.
 
The new fields $(\f^a,\psi)$ display non-standard dynamics in that they feature linear 'velocity coordinates' rather than the standard
'velocity squared' kinetic terms. It may be relevant to stress that $(\f^a,\psi)$ can be seen as 'gravitational coordinates'
 rather than
usual matter fields since on the level of the HCS densities from which the HCSG result, the Higgs scalar is on the same footing as the non Abelian
gauge connection.

The present work, which is a continuation of that done in Ref. \cite{Radu:2018fda}
for the the lowest
   dimension $d=3$, provides the explicit expression of the HCS Lagrangians up to $d=7$,
	together with an investigation  of the simplest solutions for $d=3,5$.
	These solutions have in common 
	the property that they do not backreact on the spacetime geometry, $i.e.$
their {\it effective} energy-momentum tensor vanishes.
	However, while for $d=3$ this includes the case of BTZ BH, for $d=5$
	only a maximally symmetric AdS background is allowed.
	We attribute this feature to the fact that the BTZ BH
	possesses the same amount of symmetries as pure AdS$_3$, being 
	a global identification of it 
	\cite{Banados:1992wn},
	\cite{Banados:1992gq}. On the other hand, the case of $d>3$ BHs in CSG
	are different;
	although their line-element is still BTZ-like \cite{Tchrakian:2019iem}, they are less symmetric than the AdS$_d$
	background.
	
 Finally, for $d=3$, the Ref. \cite{Radu:2018fda} has provided (numerical) evidence for the existence
of BTZ-like BH with standard asymptotics also for the fields $(\phi^a,\psi)$,
provided the action is supplemented with a Maxwell field.
We conjecture that a similar property holds in the higher dimensional case. 
In this respect, it may be interesting to consider the HCSG systems in the presence of non-Abelian matter (in $d>3$), or Skyrme scalars,
to search for regular solutions.

 \medskip
 \medskip
\noindent {\large\bf Acknowledgements}
\\
We are grateful to Ruben Manvelyan  for useful discussions.
The  work of E.R.  is  supported  by  the Center  for  Research  and  Development  in  Mathematics  and  Applications  (CIDMA)  
through  the Portuguese Foundation for Science and Technology (FCT - Fundacao para a Ci\^encia e a Tecnologia), 
references UIDB/04106/2020 and UIDP/04106/2020, and by national funds (OE), through FCT, I.P., 
in the scope of the framework contract foreseen in the numbers 4, 5 and 6 of the article 23,of the Decree-Law 57/2016, of August 29, changed by Law 57/2017, of July 19.  
We acknowledge support  from  the  projects  PTDC/FIS-OUT/28407/2017 
 and  CERN/FIS-PAR/0027/2019.  
 This work has further been supported by the European Union’s Horizon 2020 research and innovation (RISE) programme H2020-MSCA-RISE-2017 Grant No. FunFiCO-777740.  
The authors would like to acknowledge networking support by the COST Action CA16104.

\begin{small}

\end{small}

\end{document}